\documentclass[12pt]{article}
\usepackage{amsmath}
\usepackage{amssymb}
\usepackage{amsxtra}

\newcommand {\CalN} {\mathcal N}

\newcommand {\BR}   {\mathbb R}

\newcommand {\ve}  {\varepsilon}
\newcommand {\ep}  {\epsilon}

\newcommand{\g}{\mathfrak{g}}

\DeclareMathOperator{\tr} {tr}

\DeclareMathOperator{\vol}{vol}

\newcommand{\SU}{SU}
\newcommand{\U}{U}
\newcommand{\SO}{SO}

\newcommand{\OSp}{OSp}

\usepackage{hyperref}
\hypersetup{
    colorlinks = false,
    linkcolor = blue,
    anchorcolor = red,
    citecolor = blue,
    filecolor = red,
    pagecolor = red,
    urlcolor = blue
}

\usepackage[numbers,sort&compress]{natbib}
\usepackage{hypernat}

\newcommand{\ZL}{Z_{\text{1-loop}}}
\newcommand{\ZN}{Z_{\text{inst}}}

\newcommand{\preprint}[1]{\begin{table}[t]  
      \begin{flushright}               
      {#1}                             
      \end{flushright}                 
      \end{table}}                     
\renewcommand{\title}[1]{\vbox{\center\Large{#1}}\vspace{5mm}}
\renewcommand{\author}[1]{\vbox{\center\large#1}\vspace{5mm}}

\newcommand{\mN}{m_{\mathrm N}}

\begin{document}
\bibliographystyle{utphys}
\preprint{ITEP-TH-04/10}
\begin{center}
\title{
On the instantons and the hypermultiplet mass of 
\\
$\CalN=2^*$
  super Yang-Mills on $S^4$
}
\addvspace{3mm}
Takuya Okuda$^a$ and Vasily Pestun$^b$%
\footnote{On leave of absence from ITEP, 117218, Moscow, Russia.
}
\\
\addvspace{4mm}
{\it
$^a$\noindent
Perimeter Institute for Theoretical Physics\\
Waterloo, Ontario, N2L 2Y5, Canada 
\\
\addvspace{2mm}
$^b$\noindent
Center for the Fundamental Laws of Nature\\
Jefferson Physical Laboratory, Harvard University\\
\addvspace{1.5mm}
Cambridge, MA 02138, USA 
}
\end{center}
\begin{abstract}
\noindent
We show that the physical $\CalN=4$ super Yang-Mills theory on a four-sphere
with an arbitrary gauge group 
receives no instanton contributions,
by clarifying the relation 
between the hypermultiplet mass and the equivariant parameters
 of the mass-deformed theory preserving $\CalN=2$ supersymmetry.
The correct relation also implies that
$\CalN=4$ superconformal Yang-Mills theory 
with gauge group $SU(2)$
corresponds to Liouville theory on a torus
with the insertion of a non-trivial operator,
rather than the identity as have been claimed in the literature.
\end{abstract}
\vfill
In this note we show that
the  physical $\CalN=4$ super Yang-Mills\footnote{By this
we mean the 
four-dimensional gauge theory
uniquely defined by the choice of a gauge group and a coupling constant,
which is invariant under the maximal number ({\it i.e.} 32) superconformal 
charges.
} on $S^4$
with an arbitrary gauge group 
receives no instanton contributions.
This follows from the correction we make to
the relation between the mass 
and the equivariant parameters
 of the  mass-deformed version of the $\CalN=4$ theory
preserving ${\mathcal N}=2$ supersymmetry
(the so-called ${\mathcal N}=2^*$ theory).
These parameters
 enter in the one-loop and the instanton contributions
to the partition function
discussed in  \cite{Pestun:2007rz-v1}.
The correct relation also leads to the correspondence
of the $\CalN=4$ $SU(2)$ theory with
Liouville theory
\cite{Alday:2009aq} on a torus that has a non-trivial operator,
rather than the identity as have been
claimed in the literature, inserted at a puncture.

The correction is summarized as follows.
We follow the notation of \cite{Pestun:2007rz-v1}  unless we state otherwise.
Let $m = i m_E, m_E \in \BR$ be the notation  used
 in Sections 1-4 of the paper  to denote
the hypermultiplet mass for the $\CalN=2^{*}$ theory on $S^4$. 
The ``holomorphic'' instanton contribution
coming from the South Pole of the four-sphere with radius $r$,
discussed in Section 5 of  \cite{Pestun:2007rz-v1},
 is given by 
the Nekrasov instanton partition function of the ${\mathcal N}=2^*$
theory (the mass parameter $m_{\mathrm{N}}$ here was called $m$ by Nekrasov in \cite{Nekrasov:2002qd})
\begin{equation}
Z_{\text{inst}}^{{\mathcal N}=2^*}
(\ep_1,\ep_2,m_{\mathrm{N}},i a)   
\nonumber
\end{equation}
with the identification of the equivariant parameters $\ep_1$, $\ep_2$
and $\ep_+\equiv \ep_1+\ep_2$
\begin{equation}
 \boxed{ \ep_1 = \ep_2 = \frac{1}{r} = \frac {\ep_{+}}{2}
 \quad\text{and} \quad m_{\mathrm{N}} = m + \frac{\ep_{+}}{2} = i m_{E} + \frac{\ep_{+}}{2}}
\label{eq:identification}
\end{equation}
rather than $m_{\mathrm{N}} = m$ as was assumed in Section 5 of 
\cite{Pestun:2007rz-v1}.
A similar statement holds for the ``anti-holomorphic'' 
instanton contribution coming from the North Pole.%
\footnote{Relevance of the shift  by $\ep_+/2$
was also noticed in \cite{Shadchin:2005cc}.}

The main result of \cite{Pestun:2007rz-v1}, adapted for the $\CalN=2^*$
theory, is the formula for the expectation value
of the Wilson loop $W_R$ in the representation $R$ of the gauge group $G$ 
\begin{multline}
 Z_{S^4}^{\CalN=2^*} \langle W_R(C) \rangle_{\CalN=2^*; m}
= \\
= \frac {1} {\vol(G)} \int_{\g} [da] \, e^{-\frac { 4 \pi^2 r^2} {g^{2}_{YM}} (a,a) } 
Z_{\text{1-loop}}^{\CalN=2^*}(ia;m)|\ZN^{\CalN=2^*}(r^{-1},r^{-1},m+r^{-1},ia)|^2 \tr_R e^{2\pi r i a}\,,
\nonumber
\end{multline}
where $Z_{S^4}^{\CalN=2^*}$ is the full partition function and
$\langle W_R(C)\rangle$ is the normalized expectation value of the
half BPS Wilson loop along the equator $C$ of $S^4$.
We have also denoted by 
$\ZL^{\CalN=2^*}$ 
the one-loop 
contribution to the partition function.
The correction (\ref{eq:identification}) implies, as we show below, that 
at $m=0$ we obtain not just  $\ZL^{\CalN=2^*}=1$  but also 
$\ZN^{\CalN=2^*}=1$ for any gauge group $G$.
At this value of $m$ the full 32 superconformal symmetries are restored,
{\it i.e.}, the $\CalN=2^*$ theory 
is promoted to $\CalN=4$ super Yang-Mills.
Hence we have
\begin{equation}
\label{eq:ZN4}
\boxed{
\ZN^{\CalN=4}(r^{-1},r^{-1},ia)
=1
}
\end{equation}
and the complete partition
function of the $\CalN=4$ theory as well as the Wilson loop
expectation value
are given simply by the action
induced from the tree level, that is 
by the  Hermitian Gaussian matrix model \cite{Erickson:2000af,Drukker:2000rr}.
Besides several issues concerning Liouville theory discussed later,
 this  also resolves the discrepancy with the 
 localization computation of \cite{Pestun:2009nn}, where it was shown that
the $\CalN=4$ theory
reduces to the perturbative two-dimensional Yang-Mills, 
and then to the Hermitian Gaussian
matrix model. 
The localization to the two-dimensional theory in \cite{Pestun:2009nn} does not
reproduce the four-dimensional instanton contributions  summed into the Dedekind eta-function 
in \cite{Pestun:2007rz-v1}
for the partition function of the physical $\CalN=4$ Yang-Mills on $S^4$. 
After the correction (\ref{eq:identification}), the two localization computations of the partition function of the physical $\CalN=4$ on $S^4$ in \cite{Pestun:2009nn} and in \cite{Pestun:2007rz} agree in the conclusion
 that there are no four-dimensional instanton corrections to the partition funciton itself. 
The revised version of \cite{Pestun:2007rz} contains the
correct relation (\ref{eq:identification})
as well as further discussion of its consequences.

First we explain why the correction (\ref{eq:identification}) is needed.
Let us consider the theory of \cite{Pestun:2007rz-v1} 
in the neighbourhood of the South Pole, which we locally 
treat as a theory on $\BR^{4}$ to make the 
connection with Nekrasov's computation.
The gauge fields are $A_{\mu}$ with $\mu = 1,\dots, 4$. The two scalar
fields $(\Phi_9, \Phi_0)$  are grouped with the gauge fields\ into the $\CalN=2$ vector multiplet, and sometimes we 
use the notation $\Phi_0 =  i \Phi_{10} := i\Phi_0^{E}$.
The scalar fields of the $\CalN=2$ hypermultiplet are  $(\Phi_{5}, \dots, \Phi_{8})$.
 
We represent the Lorentz group $\SO(4)$ as $\SU(2)_{L} \times \SU(2)_{R}$.  
Our choice of the $\CalN=2$ supersymmetry subalgebra of the $\CalN=4$ theory, 
namely the splitting of the scalar fields into $(\Phi_5, \dots, \Phi_8)$  and $(\Phi_9,\Phi_{10})$, 
breaks the $\SO(6)$ R-symmetry of the $\CalN=4$ theory down to $\SO(4) \times \SO(2) = 
\SU(2)^{R}_{L} \times \SU(2)^{R}_{R} \times \U(1)^{R}$ in the notation 
of \cite{Pestun:2007rz-v1}. 
The factor $\SU(2)^{R}_L \times \U(1)^{R}$ is the classical $R$-symmetry 
of the $\CalN=2$ vector multiplet. The factor $\SU(2)^{R}_{R}$ is the flavour symmetry of the hypermultiplet.
 (Turning on the mass on $S^4$ for the hypermultiplet kills the $U(1)^R$ symmetry and breaks 
the $\CalN=2$ R-symmetry group $\SU(2)^R_L$ down to $\SO(2)$
and the flavour symmetry group $\SU(2)^R_R$ down to $\U(1)_F$, so that the global symmetry group 
of the $\CalN=2^*$ theory on $S^4$ is $\OSp(2|4)\times \U(1)_F$.)

The bosonic fields of the theory 
naturally transform under the symmetry groups as
\[ \overbrace{A_{1}, \dots A_{4}}^{ SU(2)_L \times SU(2)_R } \, \,
\overbrace{\Phi_5, \dots, \Phi_8}^{{ \SU(2)_L^R \times \SU(2)_R^R }} \, \,
\overbrace{\Phi_9,\Phi_{10}}^{{\U(1)^R}}\,.\] 
 The sixteen-component fermionic field $\Psi$ on the $\CalN=4$ 
theory in the convention of \cite{Pestun:2007rz-v1} is 
given in terms of four four-dimensional chiral spinors
as \[
\Psi = \begin{pmatrix}
\psi^L \\ \chi^R \\ \psi^R \\ \chi^L
\end{pmatrix}.
\] 
Each of these spinors $\psi^L,\chi^R,\psi^R,\chi^L$ has four
components. We summarize their transformation properties in the table:

\bigskip
\begin{center}
\begin{tabular}{|c|c|c|c|c|c|c|}
  \hline
  $\ve$ &  $\Psi$   &$\SU(2)_L$& $\SU(2)_R$ & $\SU(2)_L^R$ & $\SU(2)_R^R$ & $\U(1)^R$ \\
  $ * $ & $\psi^{L}$ & $1/2$ & 0  & $1/2$  & 0 & $+1/2$ \\
  $ 0 $ & $\chi^{R}$ & $0$  & $1/2$& 0   & $1/2$ & $+1/2$ \\
  $ * $ & $\psi^{R}$ & $0$  & $1/2$& $1/2$ &0 & $-1/2$ \\
  $ 0 $ &  $\chi^{L}$ & $1/2$ & 0 & 0   & $1/2$ & $-1/2$ \\
  \hline
        & $A_{1}\dots A_{4}$     & 1/2 & 1/2 & 0   &   0   &  0   \\
        & $\Phi_{5}\dots \Phi_{8}$& 0  &  0  & 1/2 & 1/2   &  0 \\
        & $\Phi_{9}, \Phi_{10}$ &   0  &  0  &  0  &  0    &  +1  \\
  \hline
        & parameters in \cite{Pestun:2007rz-v1}           &   0  & $\ep_{+}$ &  $\ep_{+}$ &  $2m$ & \\
        & parameters in \cite{Nekrasov:2002qd,Nekrasov:2003rj} & $\ep_{-}$& 
$\ep_{+}$& $\ep_{+}$ & $2 \mN - \ep_{+}$ &  \\ 
\hline
\end{tabular}
\end{center}
\bigskip
We went ahead and presented the relation between values of the equivariant parameters 
in \cite{Pestun:2007rz-v1} and \cite{Nekrasov:2002qd,Nekrasov:2003rj}
at the bottom of the table, which we are going to explain now.

Let the spinor $\ve(x)$ be the parameter of the supersymmetry transformations
(not to be confused with equivariant parameters),
and let $\ve(0)$ be the value of $\ve$ at the South Pole $x=0$.
We restrict the $\CalN=4$ supersymmetry algebra to the $\CalN=2$ subalgebra 
by taking $\ve$ in the $+1$-eigenspace of the operator $\Gamma^{5678}$. 
Such a spinor is of the form 
\[
\ve=
\begin{pmatrix}
  * \\ 0 \\ * \\ 0
\end{pmatrix},
\]
transforms in the spin-$1/2$ representation of  $\SU(2)^R_L$ and in the trivial representation of  $\SU(2)^R_R$. At the South Pole $\ve(0)$ is
of the right chirality, transforming non-trivially 
under  $\SU(2)_{R}$ and the $\CalN=2$ R-symmetry $\SU(2)^{R}_{L}$.
Since equivariant rotations should keep the spinor $\ve(0)$ invariant,
the parameters of $\SU(2)_R$ and $\SU(2)_{L}^R$ must be equal, so that their 
action on $\ve(0)$ is cancelled. These parameters are
 denoted as $\ep_{+}$ in 
\cite{Nekrasov:2002qd}. 

The equivariant 
parameters for the spatial rotation $\SU(2)_L$ and
the flavour rotation $\SU(2)^{R}_{R}$ for 
the Nekrasov's $\CalN=2$ deformed theory in the $\Omega$-background on
$\BR^4$ 
in \cite{Nekrasov:2002qd} do not have to be related to each other. 
The parameter for $\SU(2)_L$ in \cite{Nekrasov:2002qd} is called
$\ep_-$, 
and the mass parameter for $\SU(2)^{R}_{R}$ can be read off from 
the formula for 
the $k$-instanton contribution to the Nekrasov partition function 
$\ZN^{\CalN=2^*}=\sum_{k\geq 0} q^k Z_k$
\cite{Nekrasov:2002qd,Shadchin:2005mx}
\begin{eqnarray}
&&
Z_{k}
=\frac{1}{k!}
\left(
\frac{\ep_+(\mN-\ep_1)(\mN-\ep_2)}{\ep_1 \ep_2
(-\mN)(\mN-\ep_+)}
\right)^k
\nonumber\\
&&\hspace{10mm}
\times\oint
\prod_{I=1}^k
 \frac{d\phi_I}{2\pi i}
\prod_{\alpha=1}^N
\frac{
(\phi_I-\mN+\frac{1}2\epsilon_+-a_\alpha)
(\phi_I+\mN-\frac{1}2\epsilon_+-a_\alpha)
}{(\phi_I+\frac{1}2\epsilon_+-a_\alpha)
(\phi_I-\frac{1}2\epsilon_+-a_\alpha)
}
\nonumber\\
&&
\hspace{10mm}\times
\prod_{I<J}
\frac{\phi_{IJ}^2
[\phi_{IJ}^2-\ep_+^2]
[\phi_{IJ}^2-(\mN-\ep_1)^2]
[\phi_{IJ}^2-(\mN-\ep_2)^2]}
{[\phi_{IJ}^2-\ep_1^2][\phi_{IJ}^2-\ep_2^2]
[\phi_{IJ}^2-\mN^2][\phi_{IJ}^2-(\mN-\ep_+)^2]}
\label{contour}
\end{eqnarray}
This can be understood as theorem (3.7) of
\cite{Moore:1997dj} applied to
the ADHM construction for the mass deformed $\CalN=4$ theory
\cite{Dorey:1999pd}.
If the $\CalN=4$ theory were realized on $D3$-branes aligned
in the real 1234-directions, 
$k\times k$ matrices $B_1, B_2$ would describe 
the complex coordinates of $D(-1)$ instantons in the
1234-directions,
and $k\times k$ matrices $B_3,B_4$  their complex
coordinates
in the hypermultiplet scalar directions 5678.
See, for example, \cite{Bruzzo:2002xf}.

According to the theorem,
the four factors in the denominator given in the last line of (\ref{contour})
 correspond respectively to the ADHM data $B_1, B_2, B_3$ and $B_4$.
In (3.13) of \cite{Nekrasov:2002qd} 
it can be explicitly seen that $B_1$ is acted on
 by $\ep_1$ and $B_2$ is acted on
by $\ep_2$. These equivariant parameters 
correspond to the first two factors in the denominator.
The
parameters
 $(\ep_{L},\ep_{R})$
for  $\SU(2)_L \times \SU(2)_R $,
 and the ones
$(\ep_1, \ep_2)$ for $\U(1) \times \U(1)$
acting canonically on the $\BR^{4}$ represented as $\BR^{2} \oplus \BR^{2}$,
 are related as
\begin{equation}
\ep_{R} = \ep_1 + \ep_2 = \ep_{+}
,~~~\ep_{L} = \ep_1 - \ep_2  = \ep_{-}.
\label{equiv-rel1}
\end{equation}
Similarly, if we denote the parameters which act on the matrices $(B_{3}, B_{4})$ by $(\ep_3, \ep_4)$, 
and the parameters of $\SU(2)_{L}^{R} \times \SU(2)_{R}^R$ by $(\ep_L^R, \ep_R^R)$, 
they are related as 
\begin{equation}
\ep_{L}^R  = - \ep_{4} - \ep_{3},
~~ 
\ep_{R}^{R} = \ep_4 - \ep_3. 
\label{equiv-rel2}
\end{equation}
We can conclude from the last two factors in the denominator
 that $(\ep_3,\ep_4) = (-\mN, \mN - \ep_{+})$, and therefore that 
\begin{equation}
\ep_{L}^R = \ep_{+},
~~~
\ep_R^R = 2 \mN - \ep_{+}. 
\label{equiv-rel3}
\end{equation}
See also (2.5) of \cite{Bruzzo:2002xf} and (3.6) of \cite{Moore:1998et}.

As a check we note that the above relations among the equivariant 
parameters are  consistent with
the numerator in the last line of (\ref{contour})
which according to the same theorem
are associated with  the ADHM equations.
The ADHM equations for the hypermultiplet near the South Pole 
in \cite{Pestun:2007rz-v1} transform as the $\chi^L$ components, {\it i.e.},  the hypermultiplet equations are acted on by the group $\SU(2)_L \times \SU(2)^R_R$ with parameters $(\ep_L, \ep_{R}^R)$. The eigenvalues for the equivariant group action on the hypermultiplet equations are then $(\pm\frac 1 2(\ep_L + \ep_{R}^{R}), \pm \frac 1 2 (\ep_L - \ep_{R}^R))$,
which in Nekrasov's notation evaluate to $(\pm(\mN - \ep_2), \pm(-\mN + \ep_1))$
and agree with the last two factors in the numerator.

In  \cite{Pestun:2007rz-v1} 
the relation $M_{ij}M^{ij}=4m^2$ between the generators $M_{ij}$ of $\SU(2)^R_R$
and the hypermultiplet mass $m$
was used when deriving
the one-loop contribution $Z_{\text{1-loop}}^{{\mathcal N}=2^*}$.
This relation implies that the relevant equivariant parameters
are given by
$(\pm \frac 1 2 (\ep_L + \ep_{R}^{R}), \pm \frac 1 2 (\ep_L - \ep_{R}^R)) =  (\pm m, \mp m)$.  
We conclude
that the equvariant parameter $\ep_{R}^{R}$ for the  
$\SU(2)_R^{R}$ flavour symmetry group of the hypermultiplet, which
gives mass to the hypermultiplet, is  
\begin{equation}
\label{eq:m-with-flavour-generator}
  \begin{aligned}
    \ep_{R}^{R} = 2 m \quad \text{in \cite{Pestun:2007rz-v1},}
\\
\text{and}\quad  \ep_{R}^{R} = 2 m_{\text N} - \ep_{+} \quad \text{in \cite{Nekrasov:2002qd}.}
  \end{aligned}
\end{equation}
This proves the identification (\ref{eq:identification}).
We remark that in the literature on Nekrasov's partition function 
and topological strings,
the limit $\ep_1 = -\ep_2$ is often assumed, {\it i.e.}, $\ep_{+} = 0$. 
In this special case the hypermultiplet mass, if taken by definition 
as the parameter for the $\SU(2)$ flavour group, is unshifted in 
Nekrasov's notation, {\it i.e.},  $m$ and $m_{\text N}$ in (\ref{eq:m-with-flavour-generator}) 
are equal.
In the framework of \cite{Pestun:2007rz-v1}, however, 
the equivariant parameters satisfy
$\ep_1 = \ep_2$, 
hence $m$ and $m_{\text N}$ defined in (\ref{eq:m-with-flavour-generator}) are distinct.
In order to avoid confusion, one needs to be clear about
what is meant by the hypermultiplet mass for the $\CalN=2^*$ theory in the $\Omega$-background.
The equations (5.14)-(5.19) in the first version of \cite{Pestun:2007rz-v1} need to be corrected as $m \to m + \ep_{+}/2$.
It is natural to regard $m$ as the physical mass of the hypermultiplet in the $\CalN=2^{*}$ theory, 
since the $\CalN=4$ superconformal symmetry  is recovered at $m = 0$. 

Next we show that the $\CalN=4$ theory with any gauge group receives no instanton contributions,
{\it i.e.}, that $Z_k=0$ for $k\geq 1$.  Our strategy is
to exhibit,
for generic $\ep_1$ and $\ep_2$,
the supersymmetries that get restored when $ \mN=\ep_1$ or $\ep_2$
and lead to (goldstino) fermionic zero-modes in
a background with an anti-self-dual gauge field.
Since the gauge theory localizes to configurations
with such gauge fields,
$Z_k$ should vanish as $\mN \rightarrow \ep_1$ or $\ep_2$.
Then the instanton contributions
in the $\CalN=2^*$ theory on $S^4$ 
disappear
in the $\CalN=4$ limit  $m \rightarrow 0$.

To analyze the symmetries let us consider the 5-dimensional 
picture from which the 4-dimensional gauge theory 
arises via dimensional reduction.  
The partition function of the 5-dimensional theory is given by
\begin{equation}
{\rm Tr}\left[
(-1)^F e^{-\beta H} g
\right]
\nonumber
\end{equation}
with 
\begin{equation}
g=
\exp\left[
-\beta
\left(
\ep_- J_L^3
+
\ep_+ J_{R}^3 +\ep_+ 
J^{R3}_{L}  + (2\mN-\ep_+)J^{R3}_{R}
+ a
\right)\right],
\nonumber
\end{equation}
using the equivariant parameters determined
in (\ref{equiv-rel1}) and (\ref{equiv-rel3}).
We have denoted by $J_L^i, J_{R}^i, J^{Ri}_{L}$ and $J^{Ri}_{R}$
($i=1,2,3$)
the generators of the  groups
$SU(2)_L$, $SU(2)_R$, $SU(2)^R_L$ and $SU(2)^R_R$, respectively.
The spacetime is $S^1 \times \BR^4$ where the circle
has circumference $\beta$, and we
identify
fields up to symmetry transformations when going around the $S^1$.  
The definition of the $\Omega$-background involves 
Lorentz ($J^3_L, J^3_R$),  $\CalN=2$ R-symmetry ($J^{R3}_L$)
and global gauge transformations ($a$) \cite{Nekrasov:2002qd}.
For $\CalN=2^*$ 
we also perform a flavor symmetry ($J^{R3}_R$) 
transformation for the hypermultiplet.
Among the generators of the 4-dimensional
$\CalN=4$ Poincar\'e superalgebra,
only those which commute with $g$
remain symmetries in the $\Omega$-background.

Let us switch to the 4-dimensional theory by taking the $\beta\rightarrow 0$
limit.
It is again useful to view the bosonic symmetries 
$SU(2)_L\times SU(2)_R \times SU(4)^R$ 
(Lorentz $\times$ R-symmetry) of $\CalN=4$ as a subgroup of 
$Spin(10)$ and focus on its Cartan $U(1)^5$,
where each $U(1)$ rotates one factor of $\BR^2$
in $\BR^{10}=(\BR^2)^5$.  
The supercharges $Q_\alpha^A$ and $\overline Q_{A \dot\alpha}$  ($A=1,\ldots,4$)
of  $\CalN=4$ superalgebra
form a 10-dimensional chiral spinor, 
so there are an even number of $U(1)$'s 
for which
the eigenvalues 
($s_1,...,s_5$) of the generators are $+1/2$.  
The left-handed supercharges $Q_\alpha^A$
then have an even number of $+1/2$ eigenvalues for the first two (Lorentz) $U(1)$'s and an even number of $+1/2$ eigenvalues for the last three 
(R-symmetry) $U(1)$'s.
On the other hand
the right-handed supercharges $\overline Q_{A \dot \alpha}$
have an odd number of $+1/2$ eigenvalues 
for the first two as well as for the last three $U(1)$'s.

Under the combined transformation $g$, 
a supercharge changes by the phase proportional to
\begin{eqnarray}
&&\ep_1 s_1+
\ep_2 s_2+
\ep_3 s_3+
\ep_4 s_4
\nonumber\\
&=&
 \ep_1 s_1 + \ep_2 s_2 
-\mN s_3 +(\mN-\ep_+) s_4,
\nonumber
\end{eqnarray}
where we have used (\ref{equiv-rel1})-(\ref{equiv-rel3}).
First notice that the supercharges with $s_1=s_2=s_3=s_4$ are invariant.
These are the left-handed supercharges preserved by the omega-background.  Because the supersymmetry transformations generated by them involve only the self-dual part of the gauge flux, they are preserved by 
anti-self-dual instantons and do not lead to fermionic zero-modes.  
Thus in a generic omega-background such instantons can contribute.

On the other hand the supercharges with $s_1=-s_2$ 
 and $s_3=-s_4$ are 
right-handed, and are restored as $\mN$ tends to $\ep_1$ ($s_1=s_3$)
or $\ep_2$
($s_2=s_3$).  They are then broken by anti-self-dual
instantons and produce fermionic zero-modes, causing the path-integral
to vanish for $k\geq 1$.

For gauge group $G=U(N)$ a more concrete  way to see the vanishing of 
$Z_{k\geq 1}$
at $\mN=\epsilon_1$
or $\ep_2$
is to take its representation,
(3.26) of \cite{Bruzzo:2002xf}, 
in terms of the
 $N$ Young tableaux $Y_1,\ldots, Y_N$ labeling the
poles in (\ref{contour}) as well as 
 the fixed points of the equivariant
 action on the instanton moduli space.
The formula involves the horizontal
and vertical distances of the box
$(i,j)\in Y_\alpha$ to the right and the bottom edges of $Y_\alpha$,
respectively.  
The fact that any non-trivial tableau 
necessarily contains a box with vanishing horizontal
and vertical distances leads to 
$Z_k=0$ for $k\geq 1$.
We also checked explicitly that for $G=SO(N)$,
the $k=1$ contributions computed from the contour integrals
\cite{Shadchin:2005mx}
vanish as $\mN\rightarrow \ep_1$ or $\ep_2$.
We conclude therefore that for any gauge group
instanton contributions vanish when $\mN=\ep_1$ or $\ep_2$:
\begin{equation}
Z_{k\geq 1}|_{\mN=\ep_1 \text{ or } \ep_2}=0.
\nonumber
\end{equation}
Consequently $\CalN=4$ super Yang-Mills on $S^4$
receives no instanton contributions.

Let us now discuss what the identification (\ref{eq:identification}) implies
for the correspondence of $\CalN=2^*$ $SU(2)$ Yang-Mills
with Liouville theory on a torus \cite{Alday:2009aq}.
We follow the convention of \cite{Alday:2009aq}
and set the radius $r$ of $S^4$ to one
unless we note otherwise.
We also define $Q=b+1/b$.

Since the basic correspondence
is motivated by the relation between the Liouville
conformal block and the Nekrasov partition function,
the mass parameter  $m_{\text{AGT}}$ of the $\CalN=2^*$ theory, 
identified with the Liouville momentum and called $m$ in \cite{Alday:2009aq},
is equal to the mass $m_{\text N}$ used by Nekrasov in \cite{Nekrasov:2002qd}.
Thus the relation
\begin{equation}
    \boxed{m_{\text{AGT}} = m + \frac{Q}{2}}
\end{equation}
holds between $m_{\text{AGT}}$ and the mass $m$
used in \cite{Pestun:2007rz-v1} when $b=1$, $Q=2$.

As we showed above
for gauge group $U(2)$
the Nekrasov partition function becomes equal to $1$ when $\mN=\epsilon_1$
or $\epsilon_2$:
\begin{equation}
 Z^{\CalN=2^*}_{\mathrm{inst}}(\ep_1,\ep_2, \epsilon_i,i a)=1  
\text{ for }i=\text{1 or 2}\,.
\nonumber
\end{equation}
The AGT relation (3.15) of \cite{Alday:2009aq}
then implies that the conformal block for $m_{\text{AGT}}=b$ or $1/b$,
for which $\Delta_{m_{\text{AGT}}}\equiv m_{\text{AGT}}(Q-m_{\text{AGT}})=1$,
is given by ${\mathcal F}_\alpha^{~b}(q)=
{\mathcal F}_\alpha^{~1/b}(q)=1/\prod_{i=1}^\infty(1-q^i)$.
This can be independently checked using
(A.6) of \cite{Poghossian:2009mk}.%
\footnote{We thank Y. Nakayama for pointing this out.}

The one-loop contribution of the $\CalN=2^*$ theory
in \cite{Pestun:2007rz-v1} manifestly cancels at $m=0$.
To see this cancellation
from the Liouville point of view,
let us recall that the Liouville one-point function on the torus 
 is given by
\begin{eqnarray}
\langle V_{m_{\text{AGT}}}\rangle_q=\int \frac{d\alpha}{2\pi}
C(\alpha^*, m_{\text{AGT}},\alpha)|q^{\Delta_\alpha }
{\mathcal F}_{\alpha}^{~m_{\text{AGT}}}(q)|^2\,.
\nonumber
\end{eqnarray}
By using (A.14) and various other formulas in \cite{Alday:2009aq},
the DOZZ three-point function becomes
\begin{eqnarray}
C(\alpha^*, m_{\text{AGT}},\alpha)&\propto&
a^2
\left|
\frac{
\exp\left[\gamma_{b,1/b}(2a+m_{\text{AGT}}-Q)\right]
\exp\left[\gamma_{b,1/b}(2a-m_{\text{AGT}})\right]
}{
\exp\left[\gamma_{b,1/b}(2a-1/b)\right]
\exp\left[\gamma_{b,1/b}(2a-b)\right]
}
\right|^2
\nonumber\\
&=&
a^2
\left|
z_{\rm vector}^{\text{\scriptsize 1-loop}}(a)
z_{\rm adjoint}^{\text{\scriptsize 1-loop}}(a,m_{\text{AGT}})
\right|^2\,,
\label{3-pt}
\end{eqnarray}
where we have dropped the factors independent of $\alpha=Q/2+a$,
and noted that $a$ can be replaced by its complex conjugate $a^*=-a$
when we take an absolute value.
From the first line, 
we see that the one-loop contributions
cancel out and 
the 3-point function reduces
to $a^2$ when $m_{\text{AGT}}=b$ or $1/b$.
Specializing to the case $b=1$ again,
we see that we need $m_{\text{AGT}}=1$ in order for the one-loop 
factors to drop out within the integral, so that the
Hermitian Gaussian matrix model of 
\cite{Erickson:2000af,Drukker:2000rr,Pestun:2007rz-v1} is recovered.

In classical Liouville theory
as well as in  quantum Teichm\"uller theory,
the Liouville vertex operator $V_{m_{\text{AGT}}}$
with $m_{\text{AGT}}\in Q/2+ i\BR$
creates a boundary whose geodesic length
with respect to the constant curvature metric
is proportional to the imaginary part \cite{Seiberg:1990eb,Teschner:2003em}.
The $\CalN=4$ limit $m_{\text{AGT}}=1$ with $b=1$
then corresponds to a boundary of zero length, {\it i.e.},
 a puncture with deficit angle $2\pi$.

We note that the value $m_{\text{AGT}}=0$ is also special
in several ways.
The properly normalized
Liouville 
correlator is a modular form of
weights $(\Delta_{m_{\text{AGT}}},\Delta_{m_{\text{AGT}}})$ \cite{Hadasz:2009sw},
thus the gauge theory partition function is
S-duality invariant at $m_{\text{AGT}}=0$.
Also at this value there is no operator insertion
and the one-point correlator reduces to the torus partition function,
as well as the theory descends from M5-branes on a torus without 
defect operators.
We emphasize, however, that it is only at $m=m_{\text{AGT}}-1=0$
that the theory restores the full $\CalN=4$ superconformal symmetry.

Finally let us  consider the 't Hooft loop $T_{j}$ that is dual to the
Wilson loop $W_j$ in the spin-$j$ representation  
\cite{Alday:2009fs,Drukker:2009id}.
When $m_{\text{AGT}}=1$, 
the action of $T_j$
on the holomorphic conformal block 
is given, up to a phase, by  
\begin{equation}
a q^{-a^2}{\mathcal F}_\alpha^{~1}
\rightarrow
\sum_{-j\leq s\leq j}
(a-s)
q^{-(a-s)^2}{\mathcal F}_{\alpha-s}^{~1}\,,
\nonumber
\end{equation}
where the sum is over $s=-j,-j+1,\ldots,j$.
Since ${\mathcal F}_\alpha^{~1}$ does not depend on $a$,
the prescription of \cite{Alday:2009fs,Drukker:2009id}
gives 
the normalized expectation value of the 't Hooft loop as
\begin{eqnarray}
 \langle T_j \rangle
=
\frac{
\int da\,  \left(a q^{-a^2}\right)^*
\sum_{s}
\left(a-s\right)q^{-(a-s)^2}
}{
\int da\,
 \left(a q^{-a^2}\right)^* aq^{-a^2}
}
=
\sum _{s}  e^{\frac{s^2 g^2|\tau |^2}{4}  }
   \left(1+\frac{s^2 g^2 |\tau |^2}{2}\right)\,.
\label{tHooft-Liouville}
\end{eqnarray}
Let us compare it with the expectation value of the
Wilson loop 
\begin{eqnarray}
 \langle W_j \rangle
=
 \frac{
 \int da\,  \left(a q^{-a^2}\right)^*
 \sum_{s}  e^{4\pi i s a}
 a q^{-a^2}
 }{
 \int da\,
  \left(a q^{-a^2}\right)^* aq^{-a^2}
 }
=\sum _{s} e^{\frac{s^2 g^2}{4}  }
   \left(1+\frac{s^2 g^2 }{2}\right)\,.
\nonumber
\end{eqnarray}
Clearly $\langle T_j\rangle$ and $\langle W_j\rangle$ are exchanged
under the S-duality transformation $\tau\rightarrow -1/\tau$.
The final expression of (\ref{tHooft-Liouville})
agrees at weak coupling with the semi-classical result of \cite{Gomis:2009ir},
including the bubbling contributions $|s|<j$.
The same exact expression for $j=1/2$
was obtained by localizing the $\CalN=4$ theory
in the 't Hooft loop background
to instantons in the two-dimensional Yang-Mills theory
in \cite{Pestun:2009nn, Giombi:2009ek}.
It would also be interesting to see if 
the integral representation
in the middle of (\ref{tHooft-Liouville}),
which admits a wave function interpretation,
arises when the original localization technique
 of \cite{Pestun:2007rz-v1} is extended to the 't Hooft loop.

We thank Nadav Drukker, Jaume Gomis, Yu Nakayama, J\"org Teschner
and in particular  Yuji Tachikawa 
for useful discussion and correspondence.
Research at the Perimeter Institute is supported in part by the Government 
of Canada  through NSERC and by the Province of Ontario through MRI.
The research of V.P. has been partially supported 
by a Junior Fellowship from the Harvard Society of Fellows, and
grants NSh-3035.2008.2 and RFBR 07-02-00645.

\bibliography{mass-correction}
\end{document}